\documentclass{jfm}

\usepackage{graphicx}
\usepackage{natbib}
\usepackage{subcaption}
\usepackage{comment}
\ifCUPmtlplainloaded \else
  \checkfont{eurm10}
  \iffontfound
    \IfFileExists{upmath.sty}
      {\typeout{^^JFound AMS Euler Roman fonts on the system,
                   using the 'upmath' package.^^J}%
       \usepackage{upmath}}
      {\typeout{^^JFound AMS Euler Roman fonts on the system, but you
                   dont seem to have the}%
       \typeout{'upmath' package installed. JFM.cls can take advantage
                 of these fonts,^^Jif you use 'upmath' package.^^J}%
      }
  \else
  \fi
\fi
\ifCUPmtlplainloaded \else
  \checkfont{msam10}
  \iffontfound
    \IfFileExists{amssymb.sty}
      {\typeout{^^JFound AMS Symbol fonts on the system, using the
                'amssymb' package.^^J}%
       \usepackage{amssymb}%

      }{}
  \fi
\fi

\ifCUPmtlplainloaded \else
  \IfFileExists{amsbsy.sty}
    {\typeout{^^JFound the 'amsbsy' package on the system, using it.^^J}%
     \usepackage{amsbsy}}
    {\providecommand\boldsymbol[1]{\mbox{\boldmath $##1$}}}
\fi

\newsavebox{\astrutbox}
\sbox{\astrutbox}{\rule[-5pt]{0pt}{20pt}}

\title[Deformation of neutrally buoyant drops in Taylor-Couette turbulence]{Deformation and orientation statistics of neutrally buoyant sub-Kolmogorov ellipsoidal droplets in turbulent Taylor-Couette flow}

\author[Vamsi Spandan, Detlef Lohse, Roberto Verzicco]%
{Vamsi Spandan$^1$%
,\ns Detlef Lohse$^{1,3}$, and Roberto Verzicco$^{2,1}$%
\thanks{Email address for correspondence:verzicco@uniroma2.it }}

\affiliation{$^1$Physics of Fluids,University of Twente,Enschede,
PO Box 217,7500 AE ,Netherlands\\[\affilskip]
$^2$Dipartimento di Ingegneria Meccanica,University of Rome \lq Tor Vergata\rq, Via del Politecnico 1,
Rome 00133, Italy\\[\affilskip]
$^3$Max Planck Institute for Dynamics and Self-Organization, 37077, G\"ottingen, Germany
}

\pubyear{?}
\volume{?}
\pagerange{?}
\date{?; revised ?; accepted ?. - To be entered by editorial office}

\begin{document}

\maketitle

\begin{abstract}
The influence of the underlying flow topology on the shape and size of sub-Kolmogorov droplets dispersed in a turbulent flow is of considerable interest in many industrial and scientific applications. In this work we study the deformation and orientation statistics of sub-Kolmogorov droplets dispersed into a turbulent Taylor-Couette flow. Along with Direct Numerical Simulations (DNS) of the carrier phase and Lagrangian tracking of the dispersed droplets, we solve a phenomenological equation proposed by Maffettone and Minale (\emph{J. Fluid Mech.} 78, 227-241 (1998)) to track the shape evolution and orientation of approximately $10^5$ ellipsoidal droplets. By varying the capillary number $Ca$ and viscosity ratio $\hat \mu$ of the droplets we find that the droplets deform more with increasing capillary number $Ca$ and this effect is more pronounced in the boundary layer regions. This indicates that along with a capillary number effect there is also a strong correlation between spatial position and degree of deformation of the droplet. Regardless of the capillary number $Ca$, the major-axis of the ellipsoids tends to align with the stream-wise direction and the extensional strain rate eigen direction in the boundary layer region while the distribution is highly isotropic in the bulk. When the viscosity ratio between the droplet and the carrier fluid is increased we find that there is no preferential stretched axis which is due to the increased influence of rotation over stretching and relaxation. Droplets in high viscosity ratio systems are thus less deformed and oblate (disk-like) as compared to highly deformed prolate (cigar-like) droplets in low viscosity ratio systems. 
\end{abstract}

\begin{keywords}
Taylor-Couette, drops, dispersed multiphase flow, emulsions
\end{keywords}

\section{Introduction}
\label{sec:intro}

Multi-phase turbulent flows are ubiquitous in nature and several industrial applications such as cloud formation, plankton dispersion in sea, fuel sprays, emulsification, homogenisation and mixing in chemical reactors and drag reduction. In particular, two-phase flows which typically comprise a secondary phase dispersed into a primary carrier flow (e.g. droplets/bubbles dispersed into water or air) are prevalent in many applications involving drag reduction, heat transfer, liquid atomisation etc. Theoretical and numerical modelling of such flows is extremely challenging due to the wide range of length and time scales involved both in the dispersed and carrier phase. Among the many parameters that influence the global response and local statistics of two-phase systems, the shape and orientation of a non-spherical anisotropic dispersed phase plays a very important role. 

In this work, we study the shape and orientation statistics of neutrally buoyant, sub-Kolmogorov droplets dispersed into a turbulent Taylor-Couette (TC) flow. TC flow is one of the paradigmatic systems to study turbulence in shear flows (see \citet{grossman2016high} for a recent review). It is a geometrically simple system where the flow is driven by two independently rotating co-axial cylinders. The driving in a TC flow can be quantified using $Re_i=r_i\omega_i(r_o-r_i)/\nu$ and $Re_o=r_o\omega_o(r_o-r_i)/\nu$ which are the inner and outer cylinder Reynolds number, respectively ($r_i$, $r_o$ and $\omega_i$, $\omega_o$ are the radii and angular velocities of the inner and outer cylinder, respectively while $\nu$ is the kinematic viscosity of the carrier fluid). When the driving in TC flow is beyond a critical value, the flow transitions from a purely azimuthal laminar flow to a turbulent flow, during which a wide range of coherent flow patterns develop in the gap width. The various regimes that can be observed in a single phase turbulent TC flow have been summarised by \citet{andereck1986flow} for the low Reynolds number regime and  recently extended for the highly turbulent case by \citet{ostilla2014exploring}. In the last decade, two-phase TC flow has also been of interest due to the massive reduction in drag achieved with the addition of a very small concentration  of dispersed phase ($\sim O(1)$\% of bubbles in water). Moreover, through several experimental and numerical studies in turbulent channel and TC flow, it has been shown that the deformability of the bubbles is of paramount importance to achieve strong drag reduction \citep{lu2008effect,dabiri2013transition,van2013importance}. Understanding the dynamics of deformation and orientation of an anisotropic dispersed phase (for e.g. ellipsoidal bubbles or drops) and its dependence on the underlying local flow topology can thus be useful for optimising several industrial and engineering applications.  

The degree of deformation of the dispersed phase depends on many factors such as the size of the drops/bubbles, intensity of local inertial and viscous forces which tend to elongate or deform the drop from its initial spherical shape, surface tension forces which are responsible for restoring the drop back to its original spherical shape etc. \citep{rallison1984deformation,stone1994dynamics}. In his pioneering work, \citet{taylor1932viscosity,taylor1934formation} was one of the first to study theoretically the deformation of a viscous drop dispersed into a carrier fluid where he showed that the shape of a deformed drop in presence of a velocity gradient depends on the viscosity ratio of the droplet and the carrier phase $\hat \mu =\mu_d/\mu_c$ ($\mu_d$ and $\mu_c$ are the dynamic viscosities of the droplet and the carrier liquid respectively) and the capillary number $Ca=\mu_cRG/\sigma$ ($R$, $G$, and $\sigma$ are the drop radius, local shear rate, and surface tension, respectively) which is the ratio of viscous to surface tension forces. In flows where the range of velocity gradients experienced by the dispersed phase is limited and deviation of the drop shape from sphericity is small, analytical solutions to compute the shape of the drop exist \citep{chaffey1967second,cox1969deformation,frankel1970constitutive}. However, when the flow field becomes complex and deviation from sphericity is beyond a critical value, the constitutive equations become intractable.  

In turbulent flows, when the dispersed phase is larger than the Kolmogorov length scale, inertial forces play a primary role in deforming the drops which may eventually lead to breakup.  \citet{kolmogorov1949disintegration} and \citet{hinze1955fundamentals} proposed one of the earliest models to study the process of deformation and breakup of such droplets dispersed into a turbulent carrier flow. The breakup of these droplets and its distribution in homogenous isotropic turbulence was further investigated by \citet{perlekar2012droplet} using a multicomponent lattice Boltzmann method. Through Lagrangian tracking of individual droplets, these simulations showed that the local Weber number at the droplet position is strongly correlated with the degree of deformation and thus eventual breakup. A series of studies used the front-tracking approach to investigate the influence of inertial size spherical and deformable bubbles on the wall drag in a turbulent channel flow \citep{lu2005effect,lu2008effect,dabiri2013transition,tryggvason2013multiscale}. While the process of breakup and coalescence of the droplets cannot be modelled yet using this approach, it provides useful insight into the effect of deformability on the dynamics of bubbles much larger than the Kolmogorov scale ($d_b\sim100\eta_k$) in wall bounded turbulence. 

When the drops are smaller than Kolmogorov scale (sub-Kolmogorov), local viscous forces acting on the droplet are more dominant in the deformation process. In this situation \citet{cristini2003breakup} used the boundary integral formulation to study the deformation and breakup process of drops dispersed into an isotropic turbulent flow with $Re_\lambda \sim 50$ ($\lambda$ is the Taylor microscale). The process of drop deformation, elongation, neck formation and eventual breakup into satellite droplets was completely resolved and captured in these simulations. However, such fully resolved simulations come at an expense of the intensity of turbulence that can be achieved in the carrier flow field and also the number of droplets that can be simulated simultaneously. 

Due to the huge computational costs involved, simulating millions of fully resolved droplets or bubbles which can deform, breakup and coalesce in highly turbulent flow fields is still beyond reach  \citep{tryggvason2013multiscale}. However, useful insights in such flows can still be attained with the use of sub-grid models that can capture the motion and deformation of the dispersed phase. Instead of fully resolving the interaction between the dispersed phase and carrier phase, semi-empirical or phenomenological models can be used to track, deform, breakup, or coalesce a large number of droplets or bubbles in highly turbulent environments. Such an approach has been adopted by \citet{biferale2014deformation}, where they study the deformation and orientation statistics of approximately $10^4$ sub-Kolmogorov neutrally buoyant drops in homogenous isotropic turbulence. The drops were assumed to be tri-axial ellipsoids and the shape and orientation of these passively advected drops was calculated using a phenomenological model proposed by \citet{maffettone1998equation}. More recently, \citet{babler2015numerical} studied the breakup of approximately $10^5$ colloidal aggregates in canonical setups like turbulent channel flow, developing boundary layer and homogenous isotropic turbulence. A similar work by \citet{marchioli2015turbulent} investigated the statistics of ductile rupture of colloidal aggregates tracked as massless particles in a turbulent channel flow. A simple criteria for aggregate breakup or rupture was adopted by both \citet{babler2015numerical} and \citet{marchioli2015turbulent} where the occurrence of a colloid breakup is directly related to the local hydrodynamic stress being higher than a critical threshold stress. The coupling of direct numerical simulations (DNS) of the carrier phase along with semi-empirical or phenomenological models for the deformation and breakup process makes it possible to simulate millions of droplets or colloidal aggregates in highly turbulent environments. 

In this study we follow the same approach to simultaneously track approximately $10^5$ neutrally buoyant sub-Kolmogorov deformable drops in a turbulent Taylor-Couette flow. The drops are treated as passive inertialess particles while the deformation and orientation of these drops is computed based on the phenomenological model proposed by \citet{maffettone1998equation} (hereafter referred to as \lq MM\rq\ model). This approach is similar to the one adopted by \citet{biferale2014deformation} for drops dispersed in homogenous isotropic turbulence and also by \citet{de2012computational} to predict hemolysis by tracking several deformable red blood cells. 

As mentioned previously, the driving of the TC system can induce several transitions in the flow from a purely laminar state to a fully turbulent system. In this study the outer cylinder is kept stationary ($Re_o=0$) and only the inner cylinder is rotated until the system transitions into turbulence. We consider two different inner cylinder Reynolds numbers; one where there is a strong footprint of coherent structures i.e. the Taylor rolls and a higher $Re_i$ where the turbulence intensity is higher and the rolls lose its coherence. For each of these drivings, we vary the capillary number ($Ca$) and viscosity ratio ($\hat \mu$) of the droplets with the carrier fluid. Under these conditions, the questions we want to address in this paper are: Does the the degree of deformation of the droplets depend on their spatial position in the gap width? How do these drops deform in different regions of the flow like the Taylor rolls, plume ejection and plume impact regions, respectively ? \citep{ostilla2014boundary}. Are the drops more prolate or more oblate like and does the capillary number and viscosity ratio have an influence on the shape? What is the preferential direction of orientation of these drops in the boundary layers and bulk region? These are some questions we attempt to answer in this paper. 

In the following section, we elaborate on the numerical schemes and methods employed to solve the equations of the carrier phase and the dispersed phase and also give a brief review on the MM model which is used to compute the deformation and orientation characteristics of the drops. In section 3 we discuss the results of our simulations where we analyse the averaged profiles of extent of deformation of the drops, probability distribution functions (p.d.f's) of the extent of anisotropy in the dispersed phase and their orientational preferences for different $Ca$ and $\hat \mu$. Finally, in section 4 we summarise the paper and give an outlook on future line of simulations. 

\section{Governing Equations and Numerical Details}
\label{sec:num}
For the carrier flow, DNS was used to solve the Navier-Stokes equations in cylindrical coordinates using a second-order accurate finite-difference scheme with fractional time-stepping \citep{verzicco1996finite,van2015pencil}. This code has already been validated and tested extensively for a wide range of parameters to study the various spatio-temporal dynamics of high Reynolds number single phase TC turbulence for both rotating and stationary outer cylinder \citep{ostilla2013optimal,ostilla2014optimal,ostilla2014boundary,ostilla2014exploring}. The code for DNS coupled with two-way coupled Lagrangian tracking has also been validated and used previously for studying the effect of buoyant spherical bubbles on the net drag reduction in a turbulent Taylor-Couette flow \citep{spandan2016drag}.  For the purpose of this study the radius ratio and aspect ratio of the TC setup is fixed to $\eta=r_i/r_o=0.833$ and $\Gamma=L/d=4$, respectively. Periodic boundary conditions are employed in the azimuthal and axial directions and a non-uniform grid spacing (clipped Chebychev clustering) is employed only in the wall normal direction. 
Since we employ periodic boundary conditions in the axial direction, we have ensured through additional test simulations that the mean profiles and other relevant statistics discussed in section 3 are not influenced by increasing the aspect ratio \citep{ostilla2015computational,spandan2016drag}. 

The dispersed phase is composed of neutrally buoyant sub-Kolmogorov drops which are treated as passive tracers \citep{biferale2014deformation}. The dispersed phase is advected based on the local velocity of the carrier phase at the exact location of each drop. Since the dispersed phase is treated in a Lagrangian manner, the position of each droplet does not necessarily coincide with the location of grid nodes on the carrier phase mesh. This limitation is overcome with the use of a tri-linear interpolation scheme to compute the velocity of the carrier phase at the exact droplet position.  
 
\begin{equation}
\frac{d \textbf x_p}{dt}=\textbf v.
\label{eqn:ppos}
\end{equation}

The droplet position is updated after every time step according to equation \ref{eqn:ppos} using a second order Runge-Kutta scheme ($\textbf v$ is the interpolated carrier phase velocity at the droplet position and $\textbf x_p$ is the droplet position vector). In order to couple the tracking of the droplets with deformation, the droplets are assumed to be tri-axial ellipsoids. Under this assumption, the drop shape and orientation can be easily described in the form of a second-order positive-definite symmetric tensor $\pmb S$ (shape tensor). This tensor satisfies the condition $\pmb S^{-1}:\textbf {xx}=1$, where $\textbf x$ is the position vector of any point on the ellipsoid surface relative to its origin. For any known tensor $\pmb S$, the eigenvalues of the shape tensor give the square of semi-axes of the ellipsoid while the eigenvectors give the orientation of the semi-axes of the ellipsoid. In case of ellipsoidal drops, the MM model proposes a phenomenological evolution equation of the shape tensor $\pmb S$ based on the competing actions of drag which tends to elongate the droplet and surface tension which tends to recover the spherical shape corresponding to minimum surface energy. In dimensional form this equation reads:

\begin{equation}
\frac{d\pmb S}{dt} - (\pmb \Omega\cdot \pmb S-\pmb S\cdot \pmb \Omega)=-\frac{f_1}{\tau}(\pmb S-g(\pmb S)\pmb I)+f_2(\pmb E\cdot \pmb S+\pmb S \cdot \pmb E).
\label{eqn:deqn_dim}
\end{equation} 

We now non-dimensionalise as follows: $\pmb S^* = \pmb S/R^2$ where $R$ is the radius of the undistorted spherical drop. $\pmb E^*=\pmb E/G$ and $\pmb \Omega^*=\pmb \Omega/G$ are the non-dimensional strain rate and vorticity rate tensors; $\pmb E = 0.5[\nabla \pmb u + \nabla \pmb u^T ]$, $\pmb \Omega = 0.5[\nabla \pmb u - \nabla \pmb u^T]$, and $G=1/\tau_\eta$ is the inverse Kolmogorov turbulent time scale. $Ca$ is the capillary number which is the ratio of the viscous forces to the interfacial forces or can also be defined as the ratio of the interfacial relaxation time ($\tau=\mu_cR/\sigma$) to the characteristic flow time ($\tau_\eta$) i.e. $Ca=\tau/\tau_\eta$ ; $\mu_c$ is viscosity of the carrier phase and $\sigma$ the interfacial tension.

In non-dimensional form the evolution equation for the shape tensor ($\pmb S^*$) can be written as follows:

\begin{equation}
\frac{d\pmb S^*}{dt} - Ca(\pmb \Omega^*\cdot \pmb S^*-\pmb S^*\cdot \pmb \Omega^*)=-f_1(\pmb S^*-g(\pmb S^*)\pmb I)+f_2Ca(\pmb E^*\cdot \pmb S^*+\pmb S^* \cdot \pmb E^*).
\label{eqn:deqn}
\end{equation} 

Since we consider both fluids to be incompressible and no occurrence of droplet breakup or coalescence, the volume of each droplet must be conserved. This implies that the third invariant or the determinant of the tensor $\pmb S$ must be constant and the function $g(\pmb S^*)=3III_s/II_s$ embedded into the evolution equation ensures volume conservation ($II_s$ and $III_s$ are the second and third invariant of the shape tensor, respectively). The parameters $f_1$ and $f_2$ depend only on the viscosity ratio ($\hat \mu$) and are chosen to make the MM model recover the asymptotic limits for small $Ca$, viscosity ratio $1/\hat \mu \rightarrow 1$ and $\hat \mu =1$ and is given by \citet{maffettone1998equation} as follows:

\begin{equation}
f_1=\frac{40(\hat \mu+1)}{(2\hat \mu+3)(19\hat \mu+16)} \qquad f_2=\frac{5}{2\hat \mu+3},
\label{eqn:pfact}
\end{equation}   

The expression for $f_2$ under predicts the net deformation when the $Ca$ and $\hat \mu$ are large and this can be corrected by including an additional dependence of $f_2$ on $Ca$ \citep{maffettone1998equation,biferale2014deformation}. However, for the sake of brevity we use equation (\ref{eqn:pfact}) in this work. The constitutive equation for the evolution of the shape tensor (equation \ref{eqn:deqn}) has terms which would require the evaluation of the local strain rate and rotation rate tensor at the location of the droplet. An interpolation scheme similar to the one used for computing the velocity in equation (\ref{eqn:ppos}) is implemented for this purpose. Analogous to the carrier phase, azimuthal and axial periodicity is adopted for the dispersed phase, i.e. when the droplets exit an azimuthal or axial boundary they are re-entered at the opposite boundary with the same velocity, shape and orientation. The entire simulation for a single case consists of two parts. First the carrier phase is simulated without any dispersed phase until it reaches a statistically stationary state. At this point the simulation is stopped, and $10^5$ sub-Kolmogorov spherical drops are introduced in the gap width at random positions and the simulation is allowed to develop. The trajectories and the information on the shape tensor of every droplet is recorded at a frequency of approximately 20 snapshots every $\tau_\eta$ for almost 100 full inner cylinder rotations. Two different Reynolds numbers for the inner cylinder ($Re_i$=2500, 5000) are studied and in each system four different capillary numbers $Ca$ = 0.05, 0.1, 0.2 and 0.3 and two different viscosity ratios $\hat \mu$=1 and 100 are considered.  

\begin{figure}
%\centerline{\includegraphics[scale=0.5]{./figures/radprofs_1.pdf}}
\centerline{\includegraphics[scale=0.375]{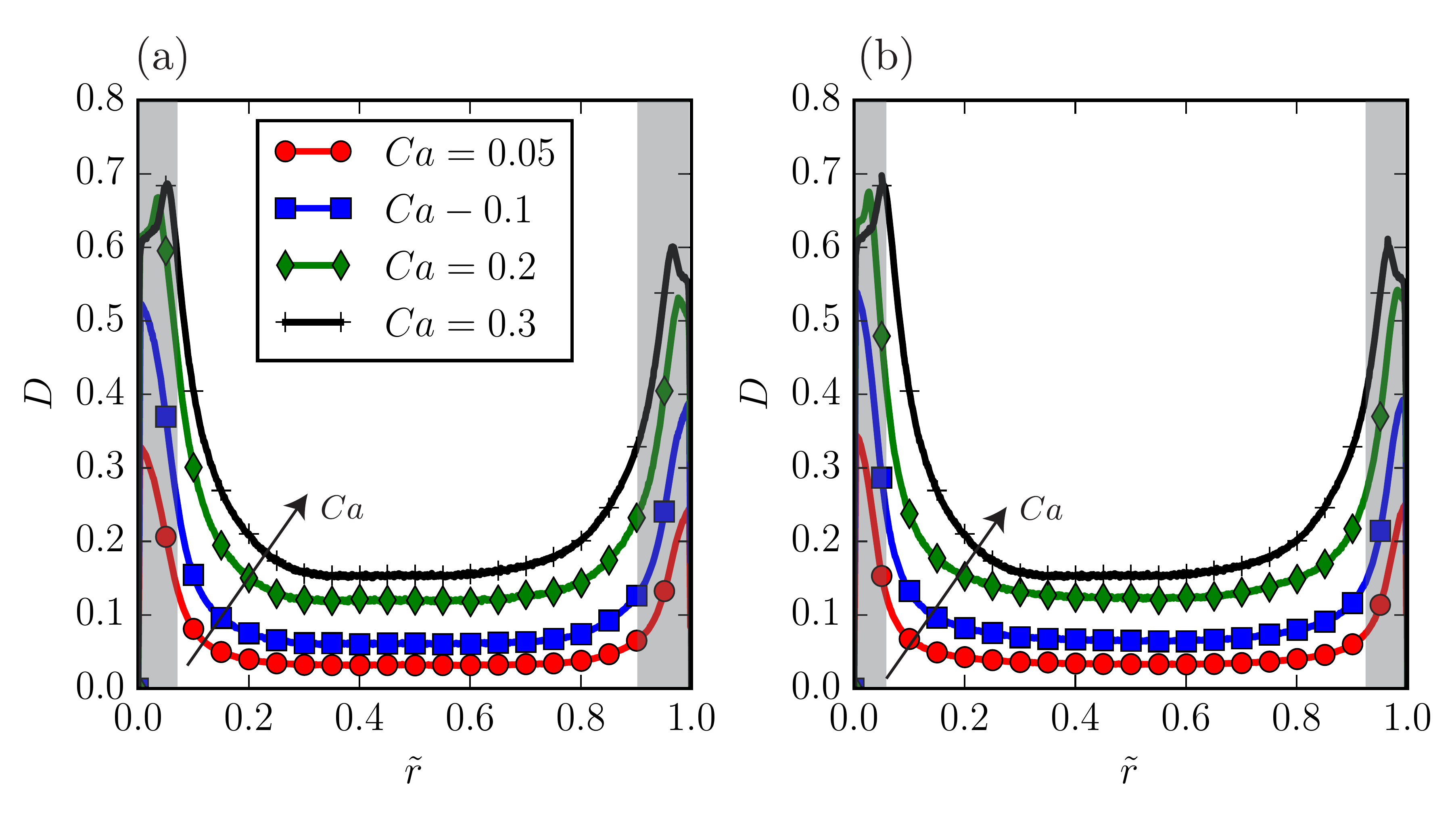}}
\caption{Azimuthally, axially and time averaged radial profiles of deformation parameters ($D$) for four different $Ca$; (a) $Re_i$=2500 (b) $Re_i$=5000. The viscosity ratio $\mu$=1. The direction of the arrow indicates increasing $Ca$. The grey shaded areas near the left and right extremes of the $\tilde r$ axis represent the inner and outer cylinder boundary layer thickness, respectively.} 
\label{fig:radprofd}
\end{figure}
  
\section{Results}
\label{sec:res}

%\begin{figure}
%\centerline{\includegraphics[scale=0.55]{./figures/axistime_2.pdf}}
%\caption{Time series of (a) three semi-axes ($d_1$, $d_2$, $d_3$) (b) normalised radial position ($\tilde r$) of a randomly chosen droplet. The temporal origin is arbitrary but after the system has reached statistical stationarity.} 
%\label{fig:timeser}
%\end{figure}

In this section we analyse the statistics of the shape and orientation of the droplets for different Reynolds $Re_i$ and capillary numbers $Ca$ while keeping the viscosity ratio fixed at $\hat \mu=1$. Since we consider the droplets to be triaxial ellipsoidal, we denote the lengths of the three semi-axes by $d_1$, $d_2$ and $d_3$ ($d_1<d_2<d_3$); i.e. $d_1$ and $d_3$ are the lengths of the minor and major semi-axes, respectively.
 
\subsection{Droplet shape distribution}

\begin{figure}
%\centerline{\includegraphics[scale=0.45]{./figures/avradd_2500_m1_inset_ibl.pdf}}
\centerline{\includegraphics[scale=0.80]{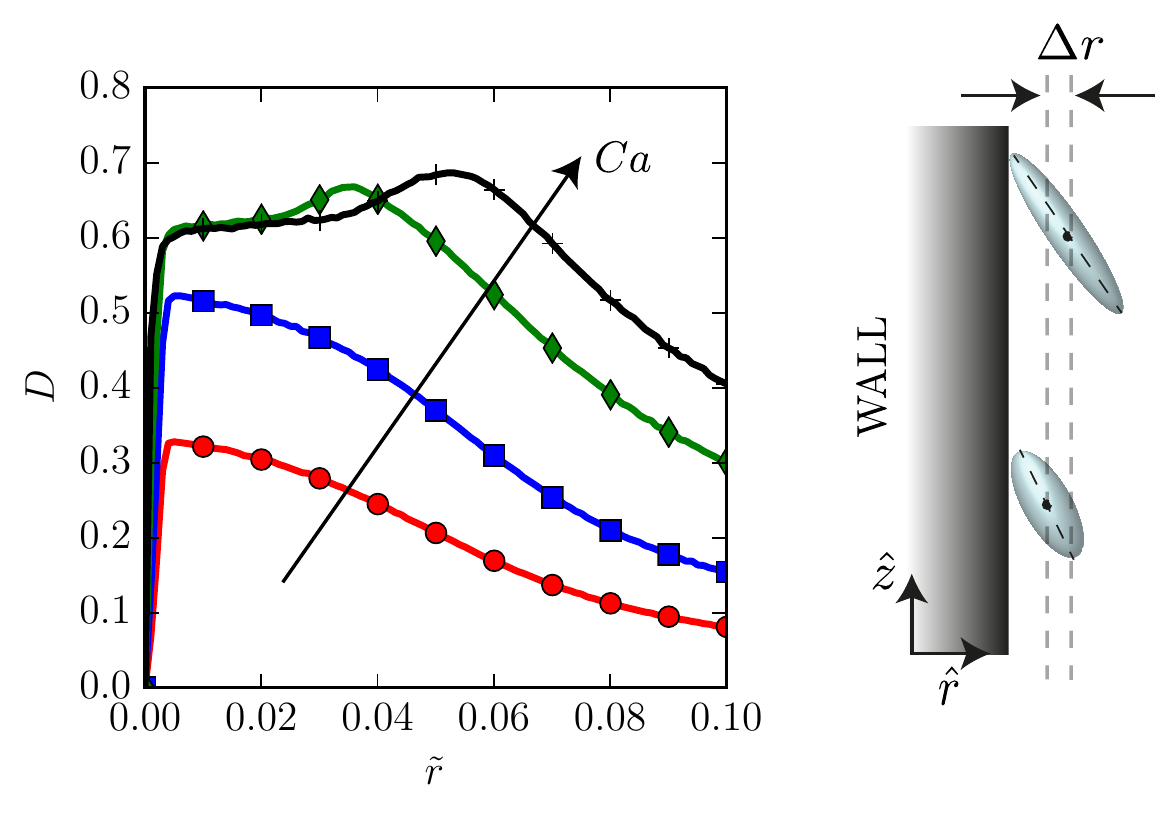}}
\caption{Zoomed in plot of figure \ref{fig:radprofd}a. The schematic on the right shows interaction of a highly deformed drop (top) and a relatively less deformed drop (bottom) with the wall. The centre of mass of a less deformed drop can approach more closer to the wall as compared to a highly deformed drop.} 
\label{fig:radprofd_in}
\end{figure}

In order to quantify the degree of deformation in the droplet shape, we make use of a non-dimensional deformation parameter defined as $D=(d_3-d_1)/(d_3+d_1)$ \citep{taylor1932viscosity,taylor1934formation,maffettone1998equation,guido1998three,guido2000drop,biferale2014deformation}. When $D\sim0$, the droplet is close to spherical, while $D\gg0$ implies the droplet is highly deformed. In order to properly quantify the correlation between the spatial position of the droplets and the degree of deformation, we plot the radial profiles of azimuthally, axially and time averaged deformation parameter for four different $Ca$ in figure \ref{fig:radprofd}. The radial position of the drops are normalised using $\tilde r=(r-r_i)/(r_o-r_i)$; the drop is close to the inner cylinder when $\tilde r \sim 0$ and close to the outer cylinder when $\tilde r \sim 1$. Along with the averaged profiles we show the inner boundary layer (IBL) and outer boundary layer (OBL) thickness using grey shaded areas which has been computed following the approach of \citet{ostilla2013optimal,ostilla2014boundary} which we describe here briefly. To compute the thickness of the IBL and OBL, a straight line is first fit through three computational nodes nearest to both walls while in the bulk a line is fit through the point of inflection and the two nearest computational nodes. The intersection of the line in the bulk along with the lines fitted near the inner and outer cylinder gives the inner boundary layer (IBL) and outer boundary layer (OBL) thickness, respectively \citep{ostilla2013optimal,spandan2016drag}. 

We can clearly observe in figure \ref{fig:radprofd} that capillary number $Ca$ has a strong influence on the mean deformation parameter $D$. As expected, owing to relatively weaker surface tension forces $D$ increases with increasing $Ca$. Additionally, as noted from figure \ref{fig:radprofd} a clear indication of the spatial dependence of the deformation parameter $D$ is visible. $D$ is higher near the boundary layers (marked by grey shaded areas) for both Reynolds numbers as compared to the bulk where it is almost three times lower. Additionally, we also observe that when $Ca=0.2$ and $Ca=0.3$, the peak of the deformation parameter $D$ lies away from the inner cylinder. To observe this deviation more clearly, in figure \ref{fig:radprofd_in} we show a zoomed-in plot of figure \ref{fig:radprofd}a and it can be clearly seen that for $Ca=0.2$ and $Ca=0.3$ there is a slight increase in $D$ away from the wall and the occurrence of this peak moves away from the wall with increasing $Ca$. This is just a consequence of the modelling of interaction of the droplet with the wall. Since we assume elastic collision between a droplet and the wall, the centre of mass of a highly deformed droplet lies farther away from the wall as compared to a less deformed droplet as shown in the schematic of figure \ref{fig:radprofd_in}). Indeed, this also depends on the relative orientation of the droplets with the wall and we will analyse this in detail in later sections.   

\begin{figure}
\centerline{\includegraphics[scale=0.4]{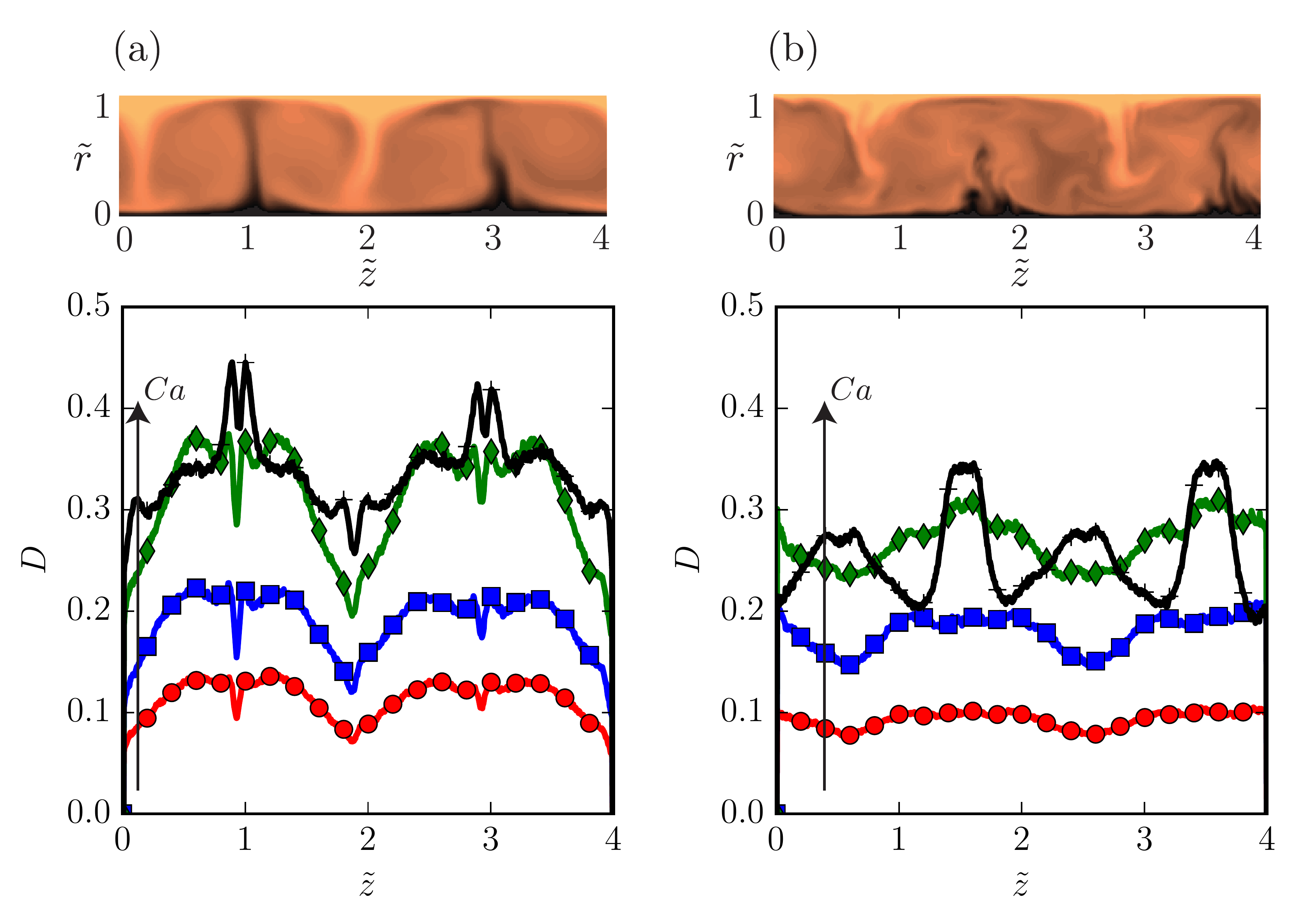}}
\caption{Azimuthally, radially and time averaged axial profiles of the deformation parameter ($D$) for four different $Ca$; (a) $Re_i$=2500 (b) $Re_i$=5000. The viscosity ratio $\mu$=1. The direction of the arrow indicates increasing $Ca$ (refer to figure \ref{fig:radprofd} for description of the markers). Instantaneous snapshots of the azimuthal velocity of the corresponding $Re_i$ are shown on the top of each panel to better visualise the plume ejections and the Taylor rolls.} 
\label{fig:axprofd}
\end{figure}

Similar to figure \ref{fig:radprofd} we plot axial profiles of the azimuthally, radially and time averaged deformation parameter $D$ as a function of the normalised axial position $\tilde z =z/d$ for $Re_i$ = 2500 and 5000 and for different $Ca$ in figure \ref{fig:axprofd}. When $Re_i$ = 2500 (c.f. figure \ref{fig:axprofd}a), we observe that regardless of the capillary number, $D$ is higher near the plume ejection regions of the inner cylinder i.e $\tilde z \sim$ 1 and 3 (also refer to the top panel in figure \ref{fig:axprofd}a which shows an instantaneous snapshot of the carrier phase azimuthal velocity). This is again a result of relatively high strain rates stretching the ellipsoidal drops in the plume ejection region near the inner cylinder; an effect likely to be produced by the Taylor rolls which make locally thinner or thicker the boundary layer produced by the cylinder rotation. A similar observation is made even for $Re_i$=5000 in figure \ref{fig:axprofd}b where the plume ejection regions near the inner cylinder are at approximately $\tilde {z}$ = 1.75 and 3.75. Although relatively high strain rates are observed in the plume ejection regions, the plumes themselves which extend from the inner cylinder into the bulk have a relatively lower strain rate owing to a strong azimuthal velocity component. This causes a sudden drop in $D$ at the exact axial position where the plumes extend into the bulk, an effect which is seen more clearly for $Re_i$=2500 due to the more coherent nature of the plumes. However, when the Reynolds number is increased the plumes are more erratic and turbulent and the sudden drop in $D$ is not clearly observed. The differences in plume structure in the bulk of the flow for both Reynolds number is clearly visible in the instantaneous snapshots of the azimuthal velocity in figure \ref{fig:axprofd}. 

\begin{figure}
\centerline{\includegraphics[scale=0.38]{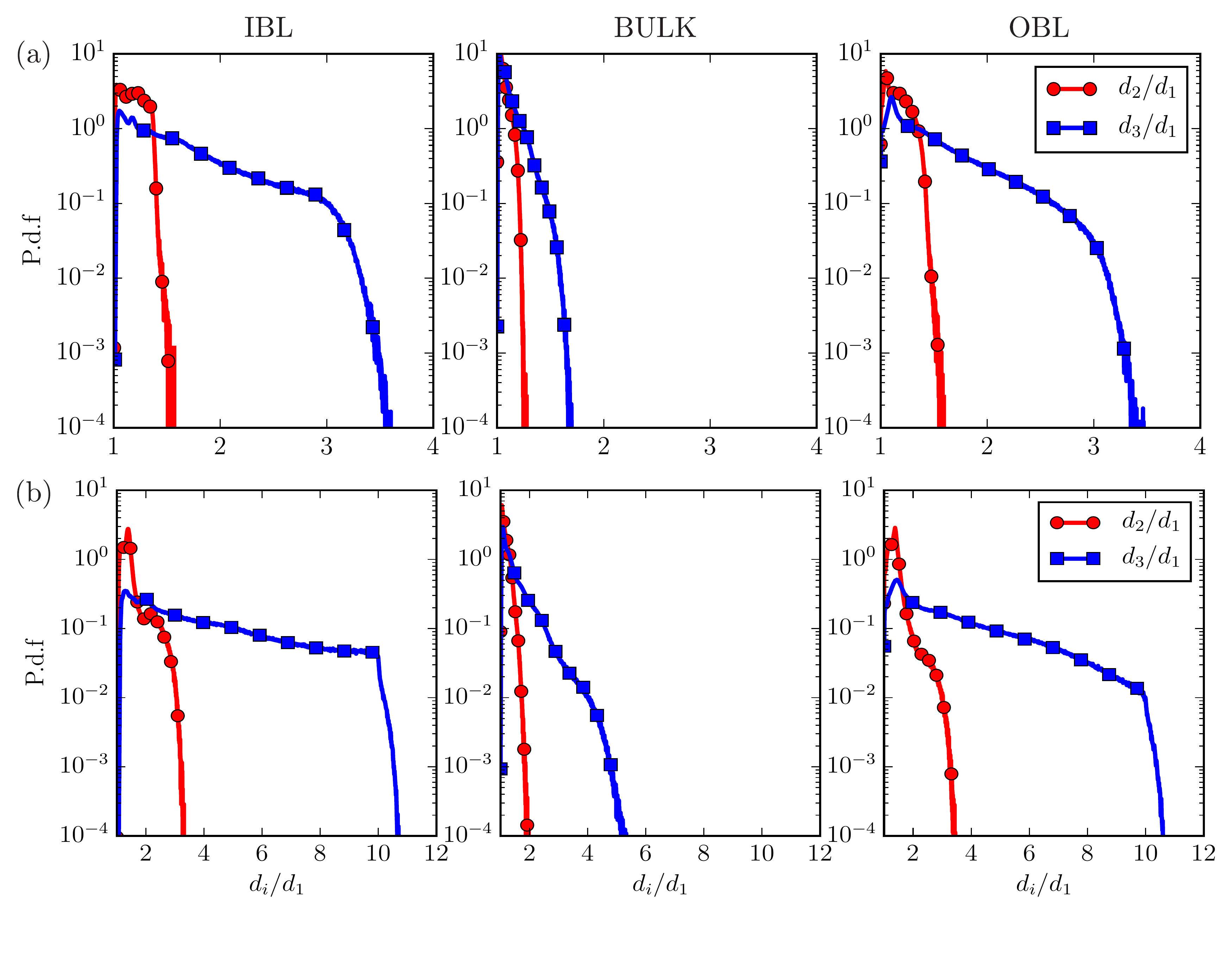}}
\caption{Probability distribution function (pdf) of the ratio of semi-axes ($d_1<d_2<d_3$) for $Re_i$=2500 and $\hat \mu=1$ (a) $Ca=0.05$ (b) $Ca=0.2$. The three panels for each $Ca$ correspond to droplets in the inner boundary layer (IBL), bulk and outer boundary layer region (OBL).} 
\label{fig:semrat}
\end{figure}

Until now we have looked at the variation of the deformation parameter $D$ with the radial and axial distance in the TC domain. Although $D$ can be used to study whether the droplet is close to spherical or highly deformed, it cannot be used to distinguish between oblate (disk-like) and prolate (cigar-like) droplets. In order to make this distinction, \citet{biferale2014deformation} used the $s^*$ parameter which was introduced by \citet{lund1994improved} to quantify the probability distribution of strain rate tensor eigen-values in turbulent flows. The shapes of the droplets can also be categorised based on simple ratios of the semi-axes i.e. $d_2/d_1$ and $d_3/d_1$ (where $d_1<d_2<d_3$). When the droplet is close to spherical $d_2/d_1 \sim d_3/d_1 \sim 1$, while for deformed droplets both $d_2/d_1>1$ and $d_3/d_1>1$. When the droplets are deformed they can additionally be categorised into oblate (disk-like) ellipsoids when $d_2/d_1 \sim d_3/d_1$ and prolate (cigar-like) ellipsoids when $d_2/d_1 \ll d_3/d_1$. 

In figure \ref{fig:semrat}, we plot the probability distribution functions (p.d.f's) of the ratio of semi-axes ($d_2/d_1$ and $d_3/d_1$) of droplets for two different capillary numbers $Ca=0.05$ and 0.2 while $Re_i=2500$ and $\hat \mu=1$. The three different panels for each capillary number correspond to the position of the droplet in the inner boundary layer (IBL), bulk region (BULK), and the outer boundary layer (OBL). When $Ca=0.05$, the droplets are almost spheroidal or in other words axis-symmetric (i.e. $d_2/d_1 \sim 1$). In the bulk region where the strain rates are relatively weaker as compared to the boundary layer region, the droplet is close to being spherical with minimum deformation (i.e. $d_2/d_1\sim d_3/d_1 \sim 1$). Additionally, the p.d.f's of $d_3/d_1$ in the IBL and OBL have much wider tails indicating that the droplets are more prolate in comparison to the bulk region.  \citet{biferale2014deformation} found that with increasing capillary number ellipsoidal drops in homogenous isotropic turbulence tend to be more cigar-like or prolate where the major axis ($d_3$) is relatively much larger than the other two axes ($d_1$, $d_2$). A similar behaviour can be observed in the bottom panels of figure \ref{fig:semrat} when the capillary number is increased to $Ca=0.2$. The tails of $d_3/d_1$ become much wider in the IBL and OBL regions as compared to $Ca=0.05$ which is a consequence of the weaker surface tension forces for higher capillary number drops. An additional consequence is that the droplets now tend to become less spheroidal and more tri-axial. 

\subsection{Orientational behaviour of the drops}

%\begin{figure}
%\centerline{\includegraphics[scale=0.5]{./figures/ang_tser.pdf}}
%\caption{Time series of the (a) stream-wise orientation of the semi-major axis (b) normalised radial position of the droplet ($\tilde r$) of a randomly chosen droplet. The temporal origin is arbitrary but after the system has reached statistical stationarity.} 
%\label{fig:ang_tser}
%\end{figure}

\begin{figure}
\centerline{\includegraphics[scale=0.4]{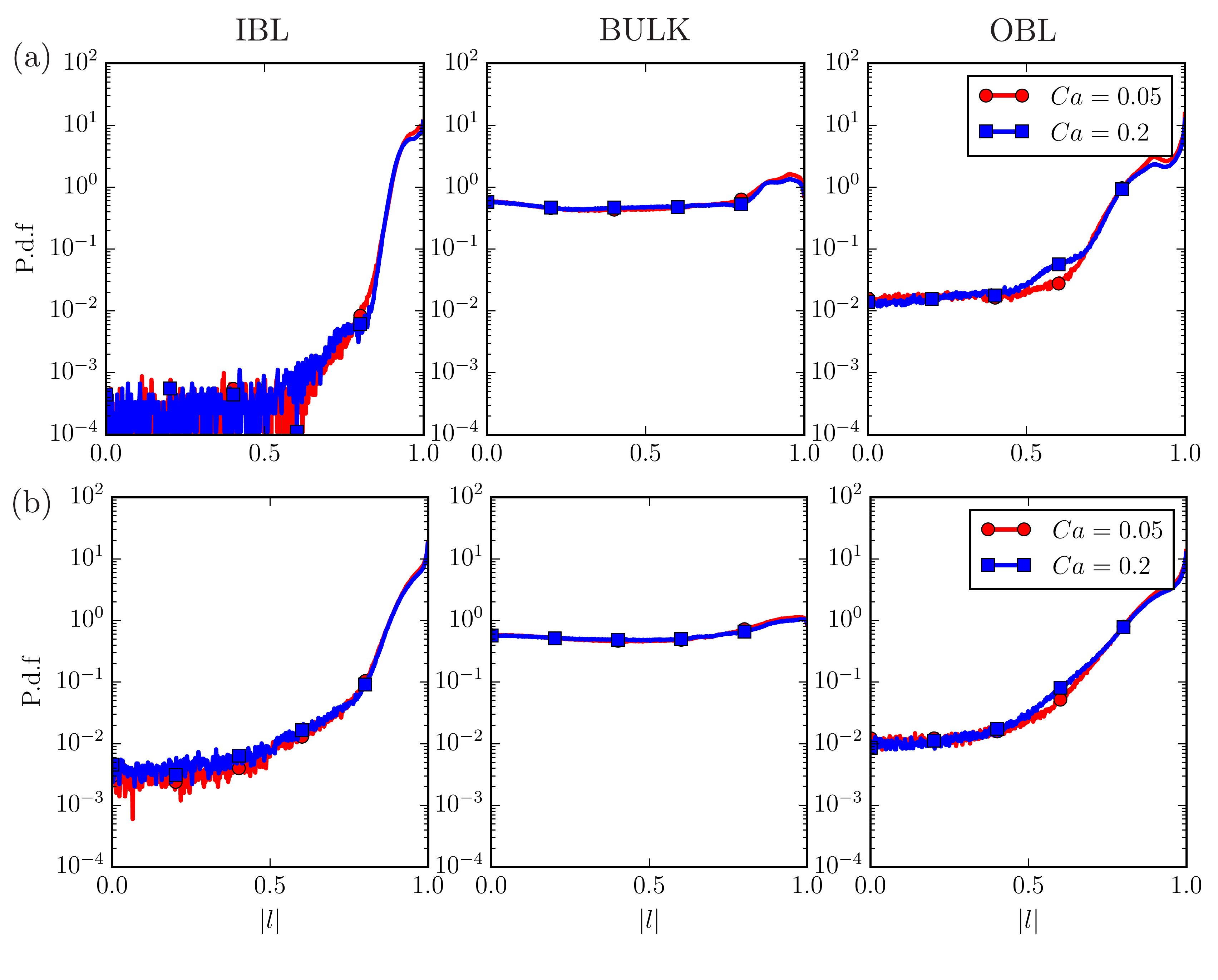}}
\caption{Probability distribution functions of the stream-wise orientation of the major axis of droplets for $Ca=0.05$ and $Ca=0.2$ (a) $Re_i=2500$ (b) $Re_i=5000$. The three panels for each $Re_i$ correspond to droplets in the inner boundary layer (IBL), bulk and outer boundary layer region (OBL).} 
\label{fig:ang_pdf}
\end{figure}

In addition to the shape and size of the ellipsoidal droplets it would also be interesting to understand their orientational preference. A number of studies \citep{mortensen2008orientation,marchioli2010orientation,njobuenwu2015dynamics} investigating the orientation behaviour of axis-symmetric solid ellipsoidal particles in wall bounded turbulent flows found that the major axis of the particles tend to align with the stream-wise direction near the wall. The shape and size of solid ellipsoidal particles does not depend on their spatial position and underlying flow conditions. However, for deforming droplets the orientation dynamics can be strongly coupled to the shape evolution which would in turn depend on the local flow conditions. \citet{biferale2014deformation} found that neutrally buoyant ellipsoidal droplets dispersed into a homogenous isotropic turbulent flow aligned with the most extensional strain-rate eigen direction and the vorticity. With an increase in the capillary number $Ca$ the alignment is lost due to an increase in the interfacial relaxation time.

\begin{figure}
\centerline{\includegraphics[scale=0.4]{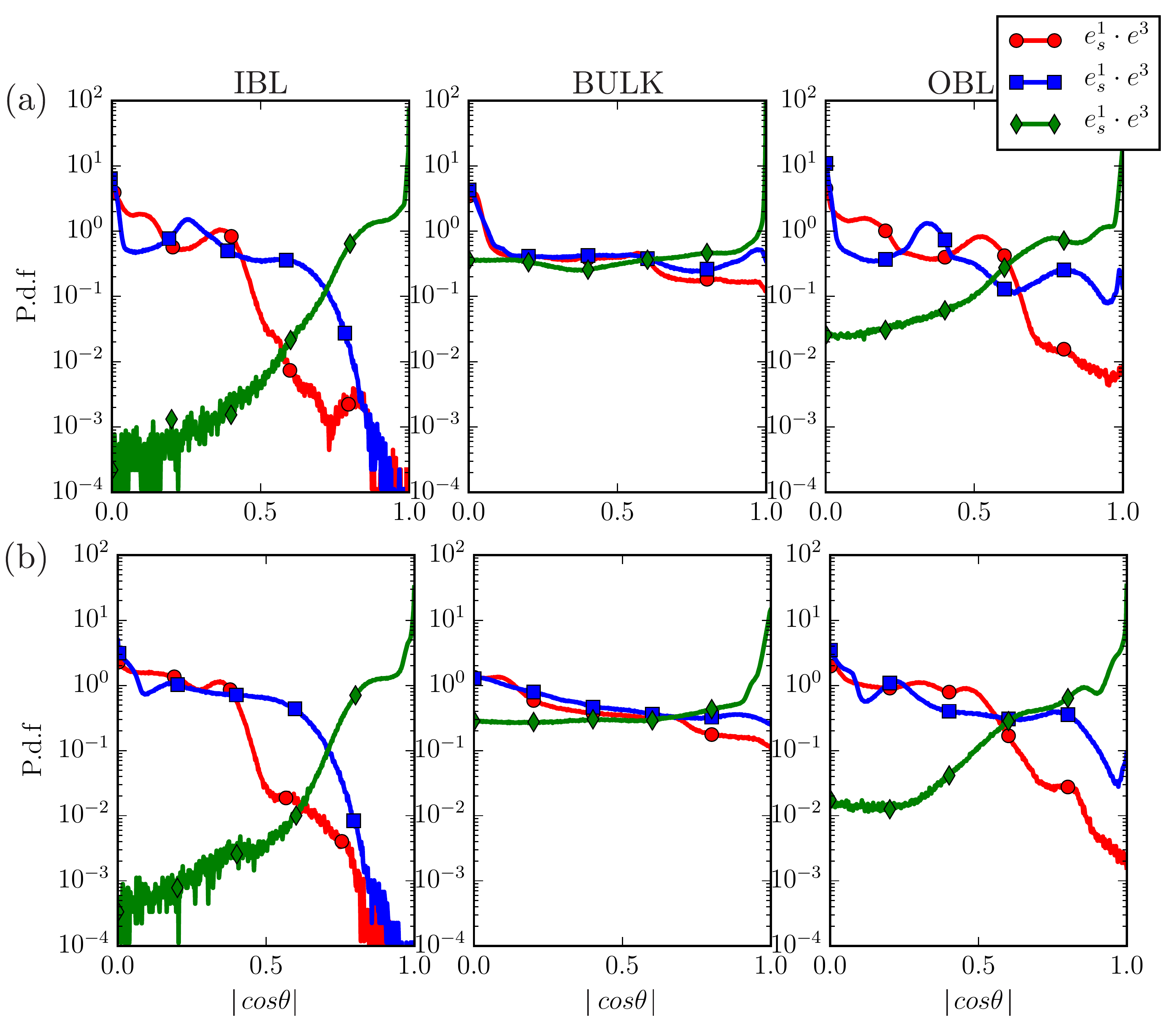}}
\caption{Probability distribution functions of the cosine of the angle between the eigen-vector of the major axis ($\bf e^3$) of the ellipsoid and three strain-rate eigen-directions ($\bf e_s^1, e_s^2, e_s^3$) $Re_i=2500$ $\hat \mu=1$ for (a) $Ca=0.05$ and (b) $Ca=0.2$.} 
\label{fig:strang_pdf}
\end{figure}

\begin{figure}
\centerline{\includegraphics[scale=0.4]{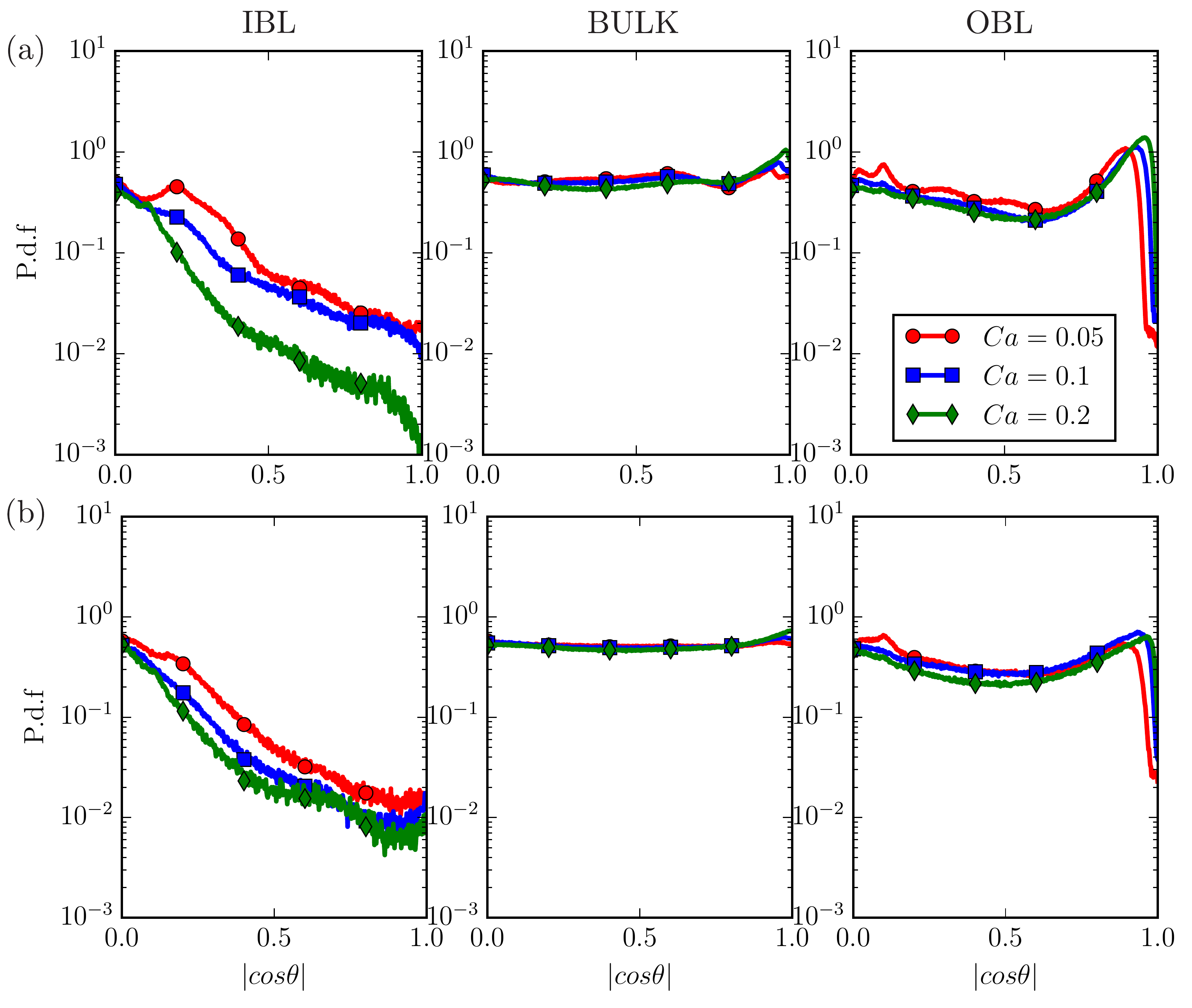}}
\caption{Probability distribution functions of the cosine of the angle between the eigen-vector of the semi-major axis ($\bf e^3$) of the ellipsoid and the vorticity ($\pmb \Omega$) $\hat \mu=1$ for (a) $Re_i=2500$ and (b) $Re_i=5000$.} 
\label{fig:vor_pdf}
\end{figure}

Here we describe the relative orientation of the major axis of the ellipsoidal droplets with respect to the coordinate system in the form of direction cosines of the angle between them i.e. $l=\mbox{cos} \theta_{\hat\theta}$, $m=\mbox{cos} \theta_{\hat r}$, $n=\mbox{cos} \theta_{\hat z}$, where $l$, $m$, $n$ are the direction cosine of the orientation with $\hat \theta$  (stream-wise), $\hat r$ (wall-normal), and $\hat z$ (span-wise) directions, respectively. In order to study the statistics of droplet orientation in detail, we plot the p.d.f's of the cosine of the angle between the droplet major axis and the stream-wise direction in figure \ref{fig:ang_pdf}. We observe that the droplets tend to be strongly aligned with the stream-wise direction (i.e. $l \sim 1.0$) in the IBL and OBL. As can be seen from the left-most panels of figure \ref{fig:ang_pdf}, the alignment is more pronounced in the case of $Re_i=2500$ as compared to $Re_i=5000$. This may be possible due to the more erratic and time dependent nature of the boundary layer region and plume ejections from the inner cylinder in the case of $Re_i=5000$ as was also seen in the contour plots of figure \ref{fig:axprofd}b. A similar behaviour is also seen in the case of solid ellipsoidal particles in the vicinity of a wall in a turbulent channel flow \citep{mortensen2008orientation,marchioli2010orientation,njobuenwu2015dynamics}. 
In the bulk, the orientation of the droplets is highly isotropic showing no specific preference in alignment with the stream-wise direction which is a consequence of the efficient mixing provided by the Taylor rolls. The capillary number $Ca$ seems to have minimal influence on the orientational behaviour of drops suggesting that it is only dominant in the deformation process and not in the effective orientation of the drops.

Now we look at the orientation statistics of the major axis with respect to the local strain-rate eigen directions and vorticity. In figure \ref{fig:strang_pdf} and figure \ref{fig:vor_pdf} we plot the pdf's of the cosine of the angle between the eigen-vector of the major axis ($\bf e^3$), the three strain-rate eigen directions ($\bf e_s^1, e_s^2, e_s^3$, where $\bf e_s^1$ and $\bf e_s^3$ are the compressive and extensional strain-rates, respectively) and vorticity ($\pmb \omega$), respectively. In homogenous isotropic turbulence, \citet{biferale2014deformation} found that the major axis of neutrally buoyant ellipsoidal droplets align with both the most extensional strain-rate eigen direction and vorticity. In figures \ref{fig:strang_pdf} and \ref{fig:vor_pdf} we find that the preference of alignment with the strain rate eigen directions and vorticity depends strongly on its relative spatial location. 

In figure \ref{fig:strang_pdf}, we consider two different configurations with capillary numbers $Ca=0.05$ and $Ca=0.2$ for a fixed Reynolds number $Re_i=2500$. In the boundary layer regions the ellipsoid major axis has a strong preference to align with the extensional strain-rate eigen direction (here the extensional strain-rate direction in the boundary layers is the azimuthal direction). In the bulk, the preference is relatively more uniform while some traces of droplets aligning with $e_s^3$ can still be found, an indication that there seems to be a competition between the extensional strain-rates trying to align itself with the droplet major axis and the homogenising nature of the Taylor rolls. This effect is clearly seen when $Ca=0.2$ given the prolate nature of the droplets (c.f. figure \ref{fig:semrat}). While the ellipsoidal droplets tend to align with the most extensional strain-rate direction in the inner boundary layer (IBL) region (left panels of figure \ref{fig:strang_pdf}) they prefer to be aligned orthogonally with the vorticity (left panel in figure \ref{fig:vor_pdf}). This is different from the behaviour of solid axis-symmetric ellipsoidal particles in an isotropic turbulent flow where rod-like particles tend to align with the vorticity \citep{shin2005rotational,parsa2012rotation,chevillard2013orientation}. The difference in behaviour is a consequence of the inhomogeneity and anisotropy of the carrier flow along with the ellipsoidal droplets continuously deforming based to the underlying flow conditions unlike solid ellipsoidal particles which are fixed in shape. With increase in Reynolds number $Re_i$, the preference of alignment with the vorticity stays qualitatively similar. Additionally, from figures \ref{fig:ang_pdf}, \ref{fig:strang_pdf} and \ref{fig:vor_pdf} we find that the capillary number $Ca$ seems to play an important role only in deforming the drops but has negligible influence on the relative orientation of the major axis of the ellipsoid while the spatial position of the droplet is crucial in the orientational preference owing to the strong inhomogeneity and anisotropy in the field. 

Until now we have looked at the effect of capillary number $Ca$ and Reynolds number $Re_i$ on the droplet deformation and orientation statistics and in the next subsection we will study the effect of viscosity ratio $\hat \mu$.

\subsection{Effect of viscosity ratio}

\begin{figure}
\centerline{\includegraphics[scale=0.41]{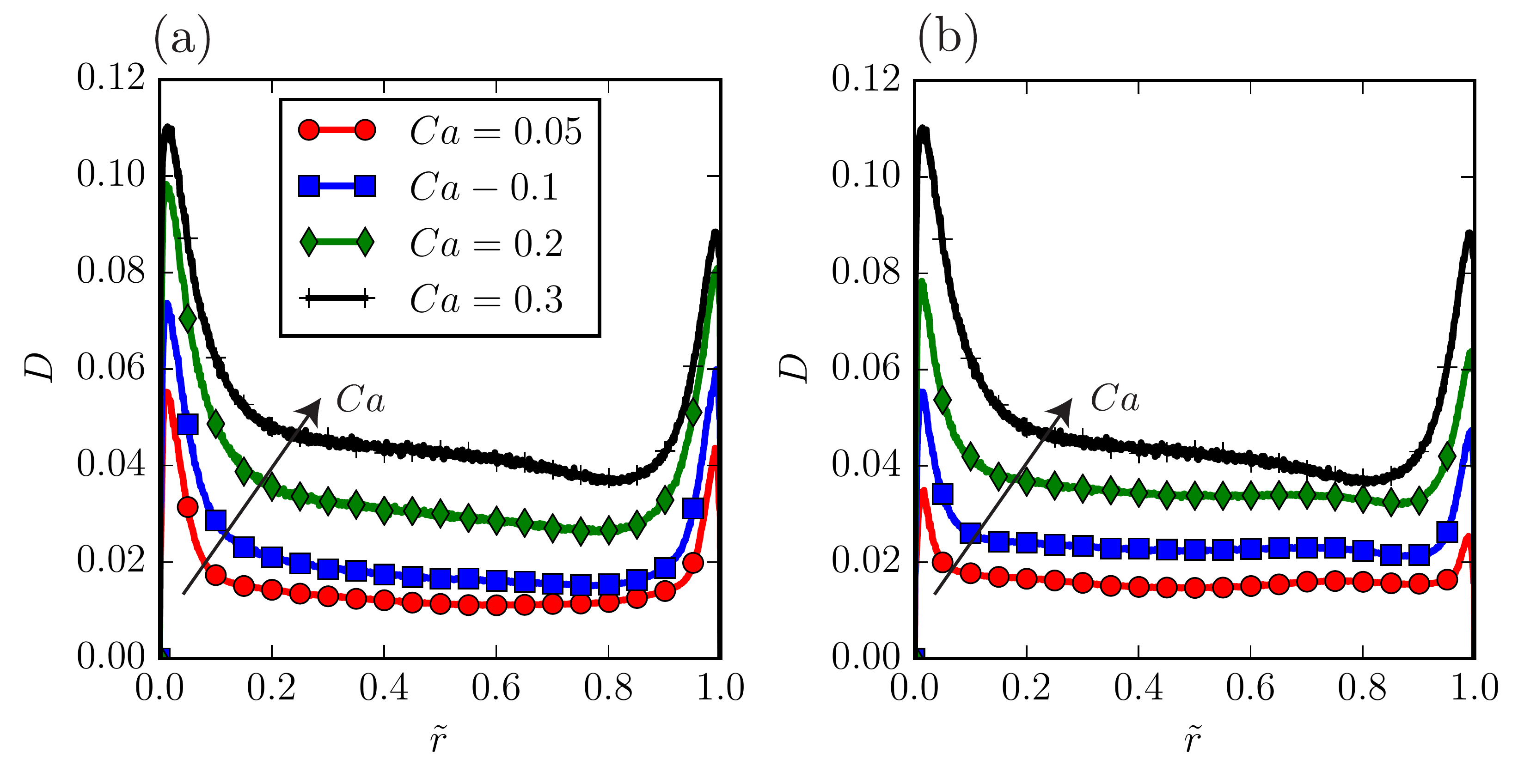}}
\caption{Azimuthally, axially and time averaged radial profiles of deformation parameters ($D$) for four different $Ca$; (a) $Re_i$=2500 (b) $Re_i$=5000. The viscosity ratio is $\hat \mu$=100. The direction of the arrow indicates increasing $Ca$. The grey shaded areas near the left and right extremes of the x-axis represent the inner and outer cylinder boundary layer thickness, respectively.} 
\label{fig:radprofd_m100}
\end{figure}

\begin{figure}
\centerline{\includegraphics[scale=0.425]{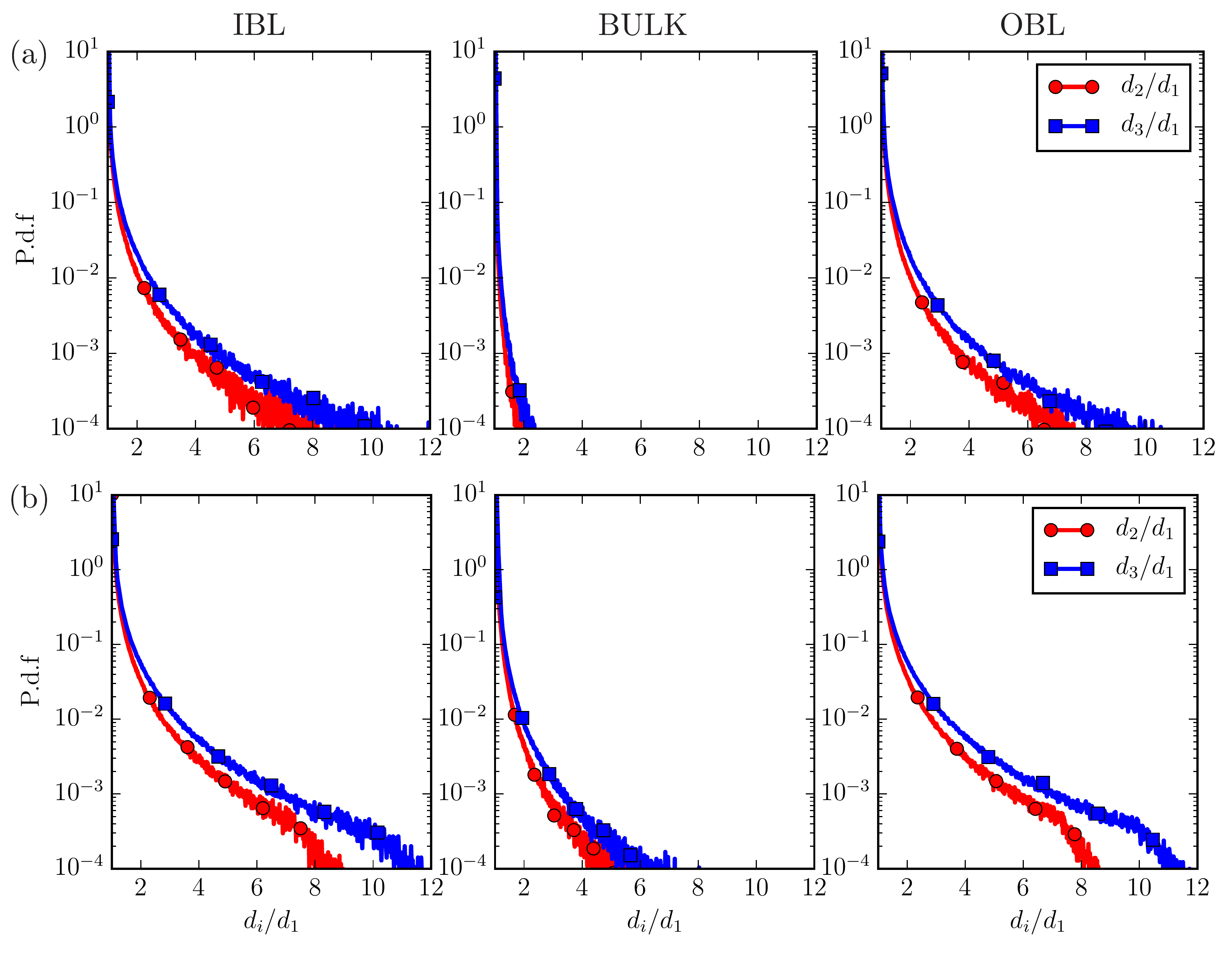}}
\caption{Probability distribution function (pdf) of the ratio of semi-axes ($d_1<d_2<d_3$) for $Re_i$=2500 and $\hat \mu=100$ (a) $Ca=0.05$ (b) $Ca=0.2$.} 
\label{fig:semrat_m100}
\end{figure}

In figure \ref{fig:radprofd_m100} we plot the averaged radial profiles of the deformation parameter $D$ for different $Ca$ in a high viscosity ratio system $\hat \mu=100$. On comparison with $\hat \mu=1$ (c.f. figure \ref{fig:radprofd}) the degree of deformation is at least 5 times smaller for all capillary numbers. This behaviour can be intuitively expected as for the same strain rates the stretching is lower for more viscous drops. Additionally, due to the relatively smaller deformations the peaks in the deformation parameter $D$ lie close to the wall for all $Ca$. While it is clear from figure \ref{fig:radprofd_m100} that there is a reduction in the overall deformation of the droplets, it would be interesting to look at the shape of the droplets as we did in figure \ref{fig:semrat}; in the case of $\hat \mu=1$ the droplets were more prolate (cigar-like). In figure \ref{fig:semrat_m100} we plot the pdf's of the ratio of semi-axes in the IBL, bulk region and OBL (similar to figure \ref{fig:semrat}). It can be seen that the semi-axes ratios $d_2/d_1 \sim d_3/d_1>1$ which is a clear indication that the droplets tend to be more oblate (disk-like) when $\hat \mu=100$. A similar behaviour is observed when the capillary number of the drops is increased. While we found previously that the capillary number plays an important role in the degree of deformation of the droplets, here we find that the droplet shape also depends strongly on the the viscosity ratio $\hat \mu$.       

\begin{figure}
\centerline{\includegraphics[scale=0.415]{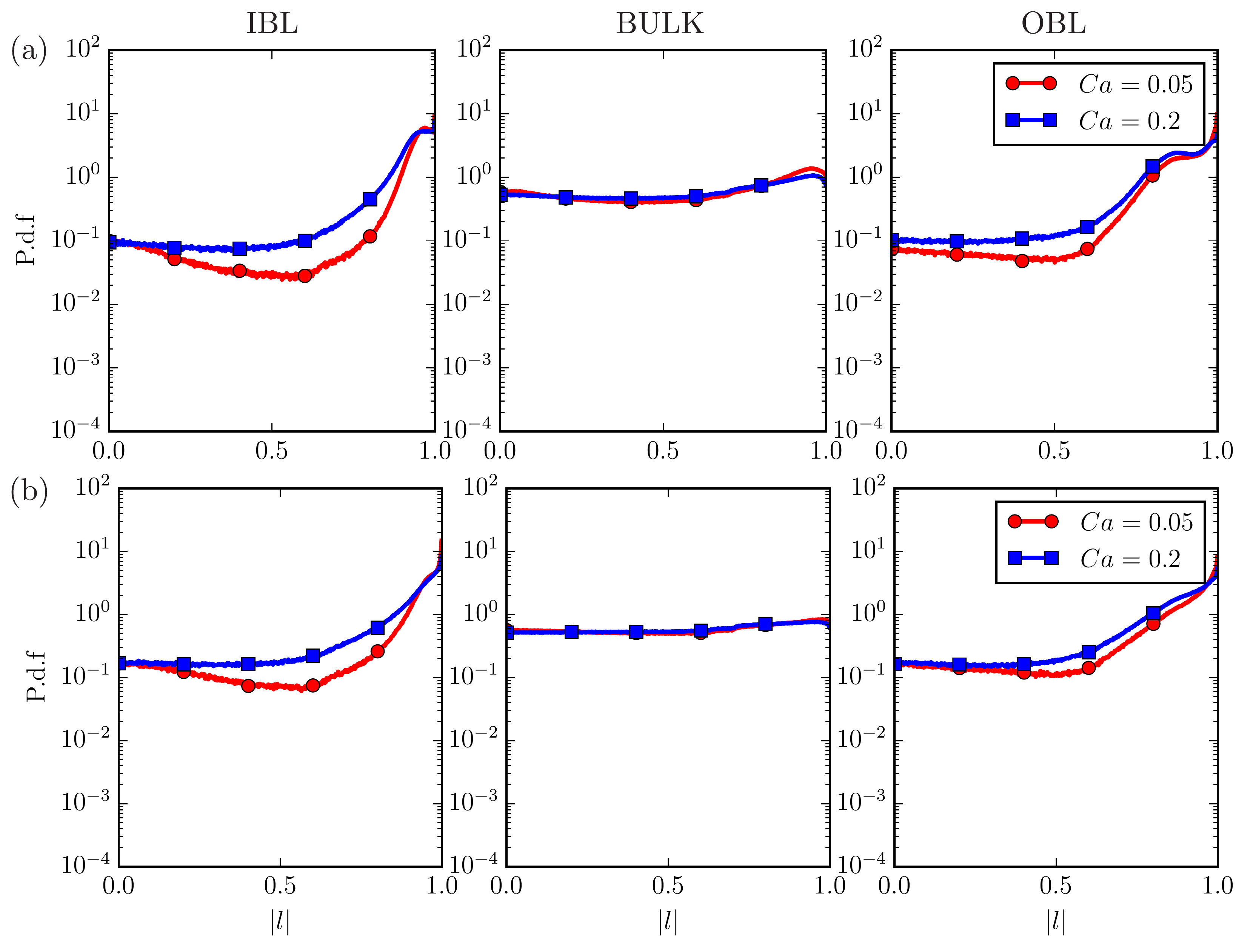}}
\caption{Probability distribution functions of the stream-wise orientation of the major axis of droplets for $Ca=0.05$ and $Ca=0.2$ and $\hat \mu=100$ (a) $Re_i=2500$ (b) $Re_i=5000$.} 
\label{fig:ang_pdf_m100}
\end{figure}

\begin{figure}
\centerline{\includegraphics[scale=0.425]{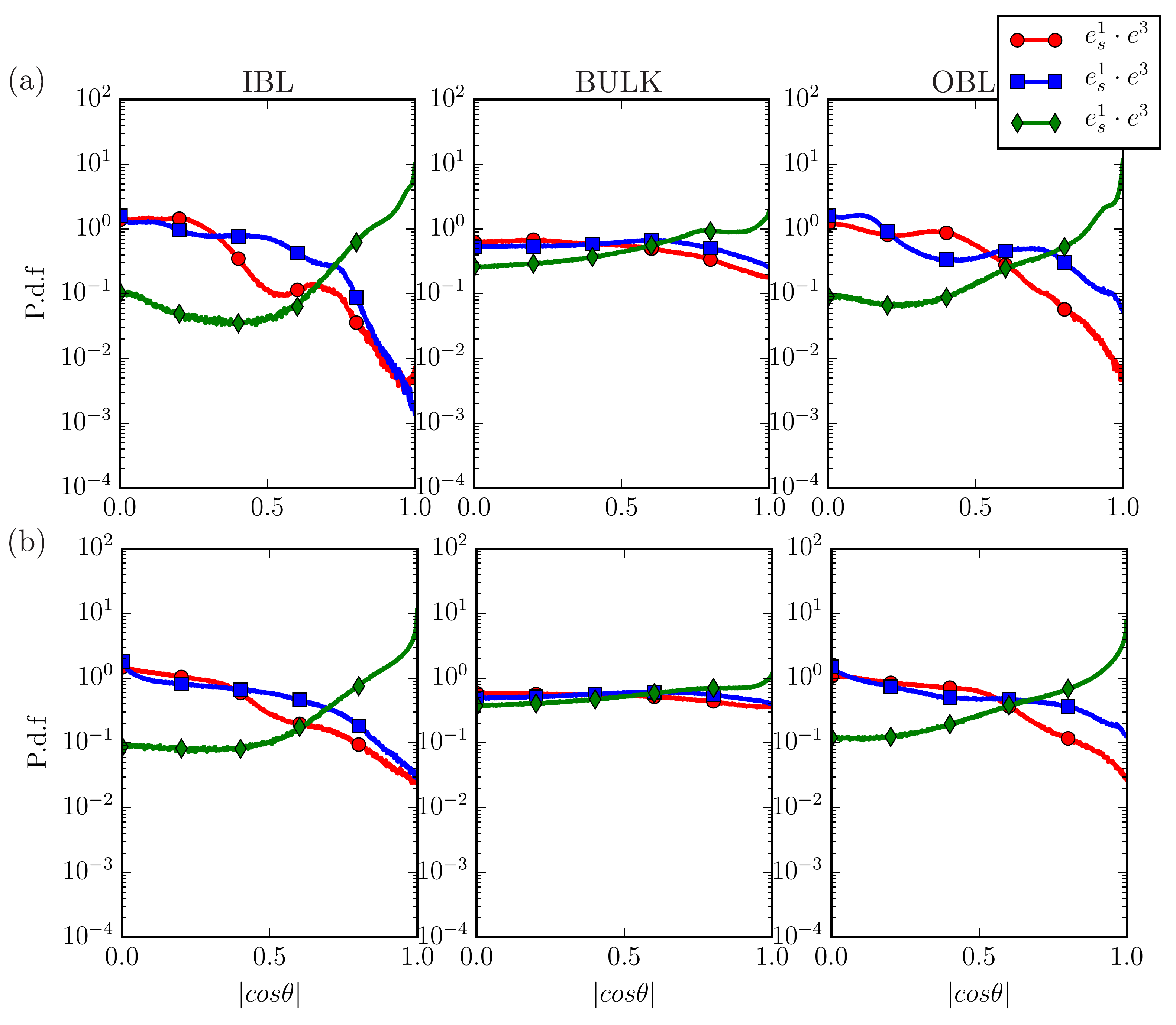}}
\caption{Probability distribution functions of the cosine of the angle between the eigen-vector of the semi-major axis ($\bf e^3$) of the ellipsoid and three strain-rate eigen-directions ($\bf e_s^1, e_s^2, e_s^3$) $Re_i=2500$ $\hat \mu=100$ for (a) $Ca=0.05$ and (b) $Ca=0.2$.} 
\label{fig:strang_pdf_m100}
\end{figure}
\begin{comment}
\begin{figure}
\centerline{\includegraphics[scale=0.415]{./figures/fig13.pdf}}
\caption{Probability distribution functions of the cosine of the angle between the eigen-vector of the semi-major axis ($\bf e^3$) of the ellipsoid and the vorticity vector ($\pmb \omega$) for $\hat \mu=100$ for (a) $Re_i=2500$ and (b) $Re_i=5000$.} 
\label{fig:vor_pdf_m100}
\end{figure}
\end{comment}

In order to understand these dependencies we need to go back to the equation governing the droplet shape (i.e. equation \ref{eqn:deqn}). By changing the viscosity ratio we influence the values of the parameters $f_1$ and $f_2$ which are calculated according to equation (\ref{eqn:pfact}). $f_1$ and $f_2$ are monotonically decreasing functions of $\hat \mu$ and while the ratio $f_1/f_2$ is almost constant (see also figure 11 of \citet{biferale2014deformation}), $f_1$ and $f_2$ decrease by up to a factor of almost 50 by increasing the viscosity ratio from $\hat \mu=1$ to $\hat \mu=100$. This has a strong influence on the effective strengths of droplet stretching, relaxation, and rotation of the ellipsoidal droplets. More importantly, in comparison to low-viscosity ratio systems the influence of rotation becomes more dominant when the viscosity ratio is high \citep{biferale2014deformation}. Additionally, reducing $f_1$ results in an increase in the effective interfacial relaxation time (i.e. $\tau/f_1$) or in other words the surface tension force acting on the ellipsoidal droplet has lesser time to adjust to the local flow conditions resulting in some highly stretched drops. The increased influence of rotation can also result in the loss of a preferential stretching axis for the ellipsoid thus resulting in more oblate or disk-like droplets. 

To understand the influence of rotation on the relative orientation we plot p.d.f's of the direction cosine of the major axis with the stream-wise direction in figure \ref{fig:ang_pdf_m100} for two different Reynolds numbers and $\hat \mu=100$. It can be immediately observed that similar to $\hat \mu=1$ the orientation is highly isotropic in the bulk region even for $\hat \mu=100$. In the IBL and OBL, while there is a tendency for the major axis to align with the stream-wise direction which is similar to what we observed for $\hat \mu=1$, the distribution of the orientation for $\hat \mu=100$ is more uniform than $\hat \mu=1$ which can be explained from earlier discussions about the increased influence of rotation. A similar behaviour is also observed in the p.d.f's of orientation of the major-axis with the eigen directions of the strain-rate tensor which are shown in figure \ref{fig:strang_pdf_m100}. In the IBL and OBL the droplets tend to align with the most extensional strain-rate eigen direction which is similar to $\hat \mu=1$ while the distribution is relatively more uniform. This is in contrast to the behaviour of oblate disk-like solid ellipsoids in an isotropic turbulent flow \citep{chevillard2013orientation} the droplets do not tend to align orthogonally to the most contracting eigen direction. These effects are connected to the anisotropy and continuously deforming nature of the drops as discussed previously. 

From figures \ref{fig:strang_pdf}, \ref{fig:vor_pdf} and \ref{fig:strang_pdf_m100} it is clear that the orientational behaviour of continuously deforming ellipsoidal droplets is completely different from that of solid axis-symmetric ellipsoids. For both viscosity ratios, the capillary number has negligible influence on the rotational behaviour of the droplets, suggesting that the droplets adjust to the underlying flow conditions regardless of the degree of deformation.  

\section{Summary and Outlook}
\label{sec:summ}

In this work we study the influence of capillary number $Ca$ and viscosity ratio $\hat \mu$ on the deformation and orientation statistics of neutrally buoyant sub-Kolmogorov drops dispersed into a turbulent Taylor-Couette (TC) flow. The drops are considered to be triaxial ellipsoids which allows us to mathematically represent the surface of the drop using a symmetric second-order positive definite tensor. By coupling the phenomenological model proposed by \citet{maffettone1998equation} for ellipsoidal drops with Lagrangian tracking and DNS for the carrier flow, we solve the time dependent evolution equation of the shape tensor of approximately $10^5$ drops dispersed into a turbulent TC flow. We find that increasing the capillary number results in large deformations of the ellipsoidal droplets due to weakening surface tension forces. Additionally, the drops deform more when they are in the inner and outer boundary layer as compared to the bulk region. Due to increased influence of rotation with increase in viscosity ratio, the droplets tend to be more oblate or disk-like as compared to more prolate or cigar-like droplets for low viscosity ratio systems. Regardless of the capillary number $Ca$ and viscosity ratio $\hat \mu$, the major axis of the drop tends to align with the stream-wise direction in the inner and outer boundary layer while the orientation is highly isotropic in the bulk; a behaviour similar to solid ellipsoidal particles in a turbulent channel flow. However, the preference of orientation with the eigen directions of strain rate tensor of these drops is very different from those exhibited by axis-symmetric ellipsoids in a isotropic turbulent flow. This is a result of the continuously deforming nature of the ellipsoidal drops dispersed into a highly inhomogeneous anisotropic flow. 

In our next line of simulations we focus on tracking continuously deforming ellipsoids with relevant hydrodynamic forces acting on them along with two-way coupling onto the carrier phase which will help us understand the influence of deformability of bubbles or drops on the dynamics of the carrier phase and consequently drag reduction. However it needs to be remembered that the bubbles/drops being sub-Kolmogorov is a limitation of the current approach and to explore the physics behind the experiments of \citet{van2013importance} where the bubbles are much larger than Kolmogorov scale, more advanced techniques such as immersed boundaries and front tracking are being implemented.

We would like to thank R. Ostilla-M{\'o}nico for various stimulating discussions during the development of the code. This work was supported by the Netherlands Center for Multiscale Catalytic Energy Conversion (MCEC), an NWO Gravitation programme funded by the Ministry of Education, Culture and Science of the government of the Netherlands and the FOM-CSER program. We acknowledge PRACE for awarding us access to FERMI based in Italy at CINECA under PRACE project number 2015133124 and NWO for granting us computational time on Cartesius cluster from the Dutch Supercomputing Consortium SURFsara.

\bibliographystyle{jfm}

\bibliography{../../mylit}

\end{document}